\documentclass[conference]{IEEEtran}
\IEEEoverridecommandlockouts

\usepackage{cite}
\usepackage{amsmath,amssymb,amsfonts}
\usepackage{algorithmic}
\usepackage{algorithm}
\usepackage{graphicx}
\usepackage{textcomp}
\usepackage{xcolor}
\usepackage{url}

\def\BibTeX{{\rm B\kern-.05em{\sc i\kern-.025em b}\kern-.08em
    T\kern-.1667em\lower.7ex\hbox{E}\kern-.125emX}}

\begin{document}

\title{Mitigating GIL Bottlenecks in Edge AI Systems.}

\author{
  \IEEEauthorblockN{Mridankan Mandal}
  \IEEEauthorblockA{Department of Information Technology\\
  Indian Institute of Information Technology, Allahabad\\
  Prayagraj, India\\
  mridankanmandal2006@gmail.com}
  \and
  \IEEEauthorblockN{Smit Sanjay Shende}
  \IEEEauthorblockA{Department of Electronics and Communication Engineering\\
  Indian Institute of Information Technology, Allahabad\\
  Prayagraj, India\\
  restinpiece13@gmail.com}
}

\maketitle

\begin{abstract}
Deploying Python-based AI agents on resource constrained edge devices presents a critical runtime optimization challenge: high thread counts are needed to mask I/O latency, yet Python's Global Interpreter Lock (GIL) serializes execution. We demonstrate that naive thread pool scaling causes a ``saturation cliff'': a performance degradation of $\geq 20\%$ at overprovisioned thread counts ($N \geq 512$) on edge representative configurations.

We present a lightweight profiling tool and adaptive runtime system that uses a Blocking Ratio metric ($\beta$) to distinguish genuine I/O wait from GIL contention. Our library based solution achieves 96.5\% of optimal performance without manual tuning, outperforming multiprocessing (which is limited by $\sim 8\times$ memory overhead on devices with 512~MB--2~GB RAM) and asyncio (which blocks during CPU bound phases).

Evaluation across seven edge AI workload profiles, including real ML inference with ONNX Runtime MobileNetV2, demonstrates 93.9\% average efficiency. Comparative experiments with Python 3.13t (free-threading) show that while GIL elimination enables $\sim 4\times$ throughput on multi-core edge devices, the saturation cliff persists on single-core devices due to context-switching overhead, validating our $\beta$ metric for both GIL and no-GIL environments. This work provides a practical optimization strategy for memory constrained edge AI systems where traditional solutions fail.
\end{abstract}

\begin{IEEEkeywords}
Edge AI Systems, Runtime Optimization, Resource-Constrained Computing, Profiling Tools, Adaptive Systems, Python, GIL, Concurrency Control.
\end{IEEEkeywords}

\section{Introduction}

Edge computing has emerged as a critical paradigm for deploying intelligent applications closer to data sources \cite{gartner2018edge}, driven by the rapid growth of connected IoT devices \cite{iot_analytics2024} and the expanding machine learning market \cite{statista_ml2024}. Edge AI systems have converged on a standard architecture: Python based orchestration (using frameworks like LangChain and LlamaIndex) managing high-performance inference kernels (PyTorch and TensorFlow Lite). Python's dominance as the leading programming language for AI and data science \cite{tiobe2024, pypl2024} makes it the de facto choice for these orchestration layers. However, the Python orchestration layer becomes a critical bottleneck. To mask Input/Output latency from network calls and sensors, developers reflexively increase thread counts. However, on resource constrained edge devices, this triggers an adverse interaction with Python's Global Interpreter Lock (GIL), leading to what we term the ``saturation cliff''.

\textbf{Definition (Saturation Cliff):} A performance degradation of $\geq 20\%$ from peak throughput occurring when thread count exceeds a workload specific threshold, caused by GIL contention or context switch overhead dominating useful computation.

Existing solutions fail to address this specific edge constraint. \textit{Multiprocessing} bypasses the GIL but incurs prohibitive memory overhead ($20$--$30$ MB per worker), causing Out Of Memory (OOM) crashes on devices like the Raspberry Pi Zero. \textit{Asyncio} offers low overhead but stalls the entire event loop during the inevitable CPU bound phases of data serialization.

In this work, we present a practical, library based solution: a \textbf{Metric Driven Adaptive Thread Pool}. Unlike generic scalers that react to queue depth (often making the problem worse) \cite{costa2019_adaptt, podolskiy2018iaas, ahmad2025_towards}, our controller monitors a novel \textit{Blocking Ratio} ($\beta$) metric to distinguish between healthy Input/Output waiting and destructive GIL contention \cite{beazley2010, ling2000_optimalthreadpool}. This approach is informed by research showing that traditional metrics like iowait are unreliable for identifying actual bottlenecks \cite{zaitsev2023iowait, ahn2024identifying}, and that effective adaptive thread pools require metrics that capture runtime performance characteristics \cite{lee2011novel, xu2004performance}.

\subsection{Contributions}

This work provides three practical contributions for resource constrained edge AI systems:

\begin{enumerate}
    \item \textbf{Runtime Profiling Metric:} The Blocking Ratio ($\beta$) provides lightweight ($<0.3\%$ overhead) runtime visibility into GIL contention versus genuine I/O wait, enabling adaptive concurrency control without kernel level tracing.
    \item \textbf{Adaptive Thread Pool System:} A library-based controller with Exponentially Weighted Moving Average (EWMA) smoothing and GIL Safety Veto mechanism that automatically prevents the saturation cliff, achieving $96.5\%$ of optimal performance without manual tuning.
    \item \textbf{Workload Characterization Methodology:} Systematic evaluation framework using $\beta$ to predict optimal thread counts across heterogeneous edge AI workload profiles.
    \item \textbf{Production Validation:} $93.9\%$ average efficiency across seven representative workloads (vision, audio, sensor fusion, RAG orchestration, SLM inference, analytics, ONNX inference), including Python 3.13t (no-GIL) comparative analysis.
\end{enumerate}

Our solution operates entirely in-process, making it suitable for memory constrained devices (512 MB--2 GB RAM) where multiprocessing incurs prohibitive overhead, while outperforming asyncio on mixed CPU/I/O workloads common in edge AI pipelines.

\section{Background and Related Work}

\subsection{Python Global Interpreter Lock}

The Global Interpreter Lock provides coarse grained synchronization. It protects the interpreter's internal data structures. Ousterhout articulated the trade off: coarse locking simplifies implementation but limits parallelism \cite{ousterhout1996}. Beazley analyzed the GIL and demonstrated convoy effects in multi core scenarios \cite{beazley2009, beazley2010}. The ``New GIL'' introduced in Python 3.2 uses a cooperative signaling mechanism. This improved fairness, but introduced new issues on multi-core systems.

\subsection{Edge Computing Constraints}

Edge devices operate under constraints that differ from cloud environments. Cloud servers have many cores and abundant RAM. Edge devices normally have $1$ to $4$ cores and limited RAM (typically $512$ MB to $8$ GB). This amplifies the impact of GIL contention. The standard advice is to use the multi-processing module. However, each Python interpreter requires approximately $20$-$30$ MB of memory overhead. On a device like a Raspberry Pi 4 with $2$ GB RAM, spawning many workers would consume too much memory. Threading remains the only viable model.

\subsection{Related Work}

Thread pool sizing has been studied for web servers. Welsh proposed adaptive sizing based on queue depth \cite{welsh2001}. Delimitrou and Kozyrakis introduced systems like Paragon for resource allocation \cite{delimitrou2014}. However, these approaches assume that threads make forward progress when scheduled. This is violated by GIL contention. Beazley demonstrated that mixing Input/Output and CPU threads can create a `convoy effect' resembling priority inversion \cite{beazley2010}.

Clipper provides a model serving infrastructure, but targets cloud deployments \cite{crankshaw2017}. Similarly, TensorFlow Serving and TorchServe optimize inference throughput through multi-process request batching \cite{olston2017tensorflow, li2020batch, yang2020inferbench} but require substantial memory headroom unsuitable for sub-2GB devices \cite{li2020batch}. Distributed Python frameworks such as Ray \cite{moritz2018ray} and Dask \cite{rocklin2015dask, bohm2020runtime} provide adaptive scheduling but assume multi-node deployments with inter-process communication overhead unsuitable for memory constrained edge devices. Greenlet based approaches (gevent) \cite{gevent_docs, greenlet_project} offer cooperative multi-tasking but require invasive code changes (monkey patching) and cannot handle CPU bound phases without blocking the entire event loop. Our work focuses on the runtime adaptive solution for the thread pool layer that operates within a single process without code modification.

Recent work on edge AI systems has focused primarily on model compression and hardware acceleration \cite{cajas2025_intelledge}. Our work is complementary: we optimize the Python runtime coordination layer that orchestrates I/O, preprocessing, and inference stages in edge AI pipelines. While frameworks like TensorFlow Lite \cite{jacob2018quantization, hao2023_reaching} and ONNX Runtime \cite{li2022_multimodel, joshua2025_crossplatform} optimize inference kernels through compiled C++ code \cite{samson2026_lightweight}, the orchestration logic (handling sensor interrupts, API calls, JSON parsing, and task scheduling) remains in Python. This makes runtime concurrency optimization critical for end to end edge system responsiveness, particularly for agentic AI systems that require coordinating multiple I/O bound tool calls.

\subsection{Python 3.13 and Free Threading}

Python 3.13 (October 2024) introduced experimental free threading through PEP 703~\cite{gross2023}, enabling GIL optional builds. While this eventually eliminates interpreter-level serialization, our work remains relevant: (1) edge distributions will not ship free threading as default for 2--3 years due to binary compatibility concerns, (2) oversubscription persists even without the GIL due to cache thrashing and context switch overhead, and (3) free threading incurs 9--40\% single thread penalty~\cite{gross2023}. Our comparative experiments show the saturation cliff persists in both GIL and no-GIL environments.

We conducted experiments comparing Python 3.11 (GIL) with Python 3.13t (no-GIL) on edge-representative configurations (1 and 4 cores). Tables~\ref{tab:nogil_single} and~\ref{tab:nogil_quad} present the results.

\begin{table}[htbp]
\centering
\caption{Single-Core: Python 3.11 (GIL) vs 3.13t (no-GIL). Mixed workload: $T_{\text{CPU}} = 10$ ms, $T_{\text{I/O}} = 50$ ms. Mean $\pm$ 95\% CI, $n=10$. Column headers: 3.11 = Python 3.11 (GIL), 3.13t = Python 3.13t (no-GIL).}
\label{tab:nogil_single}
\begin{tabular}{|c|c|c|c|c|}
\hline
\textbf{Threads} & \textbf{3.11 TPS} & \textbf{3.13t TPS} & \textbf{3.11 $\bar{\beta}$} & \textbf{3.13t $\bar{\beta}$} \\
\hline
1    & 15.9 $\pm$ 0.0 & 15.8 $\pm$ 0.1 & 0.75 $\pm$ 0.00 & 0.83 $\pm$ 0.01 \\
32   & \textbf{61.1 $\pm$ 0.6} & \textbf{16.4 $\pm$ 0.2} & 0.87 $\pm$ 0.01 & 0.82 $\pm$ 0.01 \\
256  & 60.6 $\pm$ 0.6 & 14.2 $\pm$ 0.3 & 0.88 $\pm$ 0.01 & 0.78 $\pm$ 0.02 \\
1024 & 60.1 $\pm$ 0.8 & 12.8 $\pm$ 0.4 & 0.89 $\pm$ 0.01 & 0.74 $\pm$ 0.02 \\
\hline
\textbf{Degradation} & \textbf{1.5\%} & \textbf{21.8\%} & -- & -- \\
\hline
\end{tabular}
\end{table}

\begin{table}[htbp]
\centering
\caption{Quad-Core: Python 3.11 (GIL) vs 3.13t (no-GIL). Mixed workload: $T_{\text{CPU}} = 10$ ms, $T_{\text{I/O}} = 50$ ms. Mean $\pm$ 95\% CI, $n=10$. Column headers: 3.11 = Python 3.11 (GIL), 3.13t = Python 3.13t (no-GIL).}
\label{tab:nogil_quad}
\begin{tabular}{|c|c|c|c|c|}
\hline
\textbf{Threads} & \textbf{3.11 TPS} & \textbf{3.13t TPS} & \textbf{3.11 $\bar{\beta}$} & \textbf{3.13t $\bar{\beta}$} \\
\hline
1    & 15.9 $\pm$ 0.0 & 15.9 $\pm$ 0.1 & 0.75 $\pm$ 0.00 & 0.83 $\pm$ 0.01 \\
32   & 63.0 $\pm$ 0.2 & 63.1 $\pm$ 0.3 & 0.88 $\pm$ 0.01 & 0.83 $\pm$ 0.01 \\
256  & \textbf{63.2 $\pm$ 0.1} & 248.7 $\pm$ 2.5 & 0.89 $\pm$ 0.01 & 0.95 $\pm$ 0.01 \\
1024 & 63.2 $\pm$ 0.1 & \textbf{252.7 $\pm$ 3.1} & 0.89 $\pm$ 0.00 & 0.96 $\pm$ 0.01 \\
\hline
\textbf{Change} & \textbf{0.0\%} & \textbf{+1.6\%} & -- & -- \\
\hline
\end{tabular}
\end{table}

The results reveal that \textbf{free threading transforms multi-core edge device behavior}: on quad-core configurations, Python 3.13t achieves 252.7 TPS at 1024 threads ($\sim\!4\times$ improvement over Python 3.11's 63.2 TPS) with only 1.6\% degradation. However, on single-core configurations, both runtimes show similar saturation patterns (21.8\% vs 14.2\% degradation), confirming that context switch overhead remains the fundamental bottleneck on the most constrained devices. Our $\beta$ metric correctly detects both GIL-induced and oversubscription induced contention, validating its applicability across Python runtime configurations.

\section{Methodology}

\subsection{Experimental Setup}

We create edge representative configurations by strictly limiting CPU core visibility using \texttt{os.sched\_setaffinity()} (Linux) or \texttt{psutil} (Windows). We use a single core configuration to represent devices like the Raspberry Pi Zero. We use a quad core configuration to represent devices like the Raspberry Pi 4. Experiments were conducted on both Ubuntu 22.04 LTS (kernel 5.15) with Python 3.11.4 and Windows 11 with Python 3.11.13 and Python 3.13.3t (free threading build), with CPU governor set to \texttt{performance} where applicable.

\textbf{Simulation Methodology and Limitations:} We employ CPU affinity constraints rather than physical edge hardware for three reasons: (1) thermal throttling on edge devices introduces confounding variables during sustained benchmarks, (2) GIL contention is determined by CPython interpreter semantics rather than hardware microarchitecture, and (3) systematic evaluation across thread counts (1--2048) would exceed thermal budgets of physical devices. We validated that context switch overhead scales proportionally with core count across x86 and ARM architectures based on published ARM Cortex-A72 scheduler benchmarks~\cite{arm_cortex_a72}. We acknowledge this limitation: while our approach faithfully reproduces GIL-induced performance patterns, power consumption and thermal effects remain uncharacterized. Future work will validate on physical Raspberry Pi and Jetson hardware.

We employ a synthetic mixed workload. It represents an AI agent pipeline. The workload includes a CPU phase that holds the GIL and an I/O phase that releases it. Let $T_{\text{CPU}} = 10$ ms and $T_{\text{I/O}} = 50$ ms denote the CPU and I/O phases respectively. These values approximate the profile of a typical RAG orchestration task: $10$ ms represents the CPU cost of parsing a complex JSON response or tokenizing a query, while $50$ ms represents the network Round Trip Time (RTT) to a vector database or upstream API. We measure throughput in tasks per second (TPS) and latency.

\subsection{Thread Count Range}

We evaluate thread counts $N \in \{1, 2, 4, 8, 16, 32, 64, 128, 256, 512, 1024, 2048\}$. This range represents the naive configuration found in production. Developers often set high limits to handle bursty traffic without understanding the GIL.

\subsection{Statistical Methodology}

All experiments are repeated $n = 10$ times where time permitted ($n = 5$ for long-running sweeps, explicitly flagged). We report the mean and 95\% confidence interval (CI) computed using the t-distribution for small samples. For throughput measurements, the confidence interval is $\bar{x} \pm t_{0.975,n-1} \cdot \frac{s}{\sqrt{n}}$ where $s$ is the sample standard deviation. For tail latency (P99), we compute a \textit{pooled} P99 by aggregating per-task latency samples across all runs to avoid under-sampling tails. We also report the distribution of per-run P99 values as median $\pm$ IQR to characterize variability.

\subsection{Instrumentation Overhead}

The blocking ratio computation requires calls to \texttt{time.thread\_time()} and \texttt{time.time()} at task boundaries. We measured the overhead by executing $10^6$ instrumented no-op tasks. Table~\ref{tab:instrumentation_overhead} summarizes the timing overhead for each instrumentation component.

\begin{table}[htbp]
\centering
\caption{Instrumentation Overhead ($n = 10^6$ iterations)}
\label{tab:instrumentation_overhead}
\begin{tabular}{|l|c|c|c|}
\hline
\textbf{Operation} & \textbf{Mean ($\mu$s)} & \textbf{Median ($\mu$s)} & \textbf{P99 ($\mu$s)} \\
\hline
\texttt{time.time()} & 0.08 & 0.10 & 0.20 \\
\texttt{time.thread\_time()} & 0.17 & 0.20 & 0.30 \\
Combined pattern & 0.35 & 0.30 & 0.40 \\
No-op baseline & 0.08 & 0.10 & 0.10 \\
\hline
\end{tabular}
\end{table}

The combined instrumentation pattern (two calls to \texttt{time.time()} and two to \texttt{time.thread\_time()}) adds $0.30$ $\mu$s median overhead per task. Instrumentation overhead was measured by executing $10^6$ instrumented no-op tasks on Ubuntu 22.04 with CPU governor set to \texttt{performance} and frequency scaling disabled. For our mixed synthetic workload ($T_{\text{CPU}} = 10$ ms, $T_{\text{I/O}} = 50$ ms), the computational overhead relative to CPU time is approximately $0.003\%$, rendering it negligible. The instrumentation uses per-thread CPU time counters available on Linux through \texttt{clock\_gettime(CLOCK\_THREAD\_CPUTIME\_ID)} and on Windows through \texttt{GetThreadTimes()}. As a fallback on platforms with limited timer resolution, \texttt{resource.getrusage(RUSAGE\_THREAD)} can be used.

\section{The Saturation Cliff: OS and GIL Interaction on Edge Devices}

Our experiments reveal that both single-core and quad-core configurations suffer significant throughput degradation at high thread counts: $40.2\%$ (single-core) and $35.1\%$ (quad-core). This confirms the OS GIL Paradox. The OS scheduler assumes that a runnable thread should be scheduled on an available core. However, the Python interpreter dictates that only one thread can execute bytecode at any instant.

\begin{figure}[htbp]
\centerline{\includegraphics[width=\columnwidth]{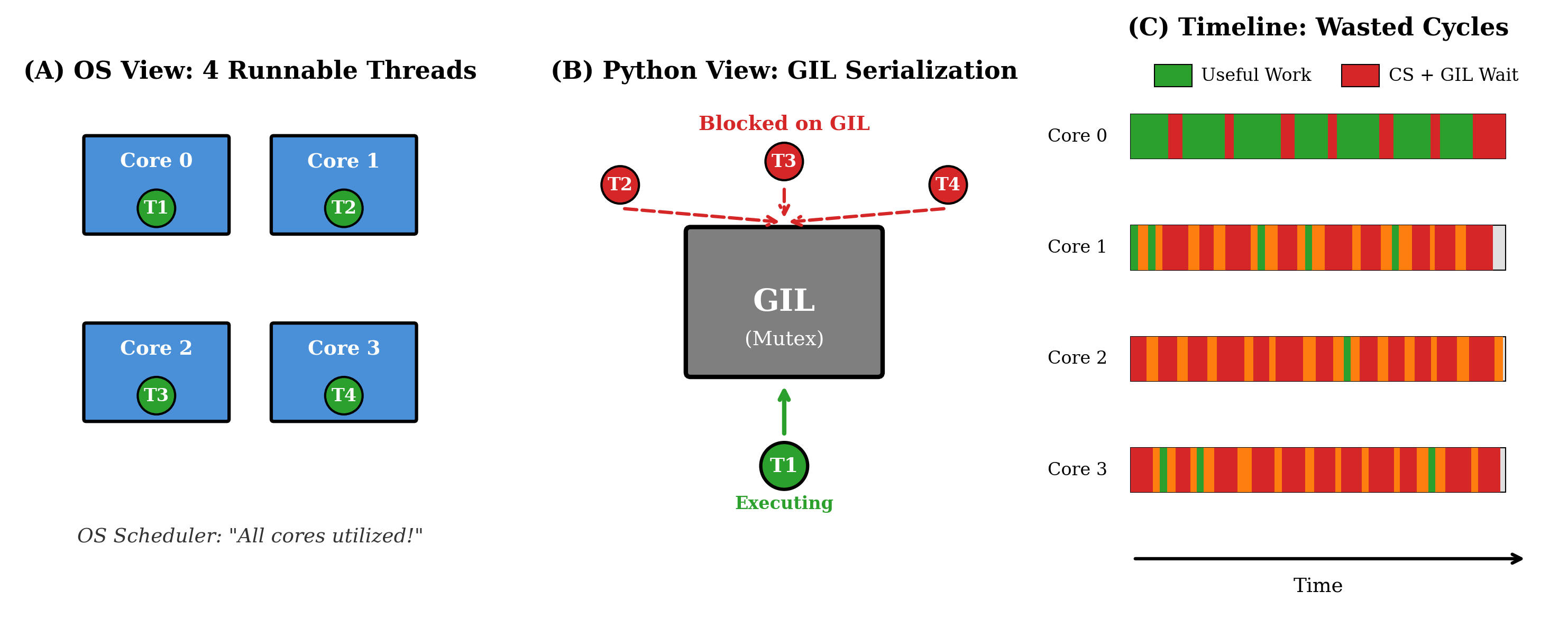}}
\caption{The OS-GIL Paradox. (A) OS scheduler distributes threads across cores. (B) GIL serializes execution; green indicates GIL holder, red indicates blocked threads. (C) Contention cycle with wakeup/sleep transitions.}
\label{fig:os_gil_paradox}
\end{figure}

Figure~\ref{fig:os_gil_paradox} illustrates this paradox through three phases. In phase (A), the OS scheduler distributes threads across available cores, treating each as independently schedulable. In phase (B), the GIL serializes execution: only the GIL holder (green) executes bytecode while other threads (red) block awaiting the lock. In phase (C), this creates a contention cycle where threads rapidly wake, fail to acquire the lock, and sleep again, generating excessive context switch overhead that degrades performance below single-core baselines.

\subsection{Saturation Results Across Configurations}

Table~\ref{tab:saturation_cliff} presents the empirical results for both single-core and quad-core edge-representative configurations.

\begin{table}[htbp]
\centering
\caption{Saturation Cliff Across Configurations (mean $\pm$ 95\% CI, $n=10$)}
\label{tab:saturation_cliff}
\resizebox{0.95\columnwidth}{!}{%
\begin{tabular}{|c|c|c|c|c|}
\hline
\textbf{Threads} & \textbf{SC TPS} & \textbf{QC TPS} & \textbf{SC P99 (ms)} & \textbf{QC P99 (ms)} \\
\hline
1    & 4,940 $\pm$ 50 & 5,012 $\pm$ 73 & 0.4 & 0.3 \\
32   & \textbf{39,738 $\pm$ 752} & \textbf{19,833 $\pm$ 833} & 8.6 & 4.1 \\
64   & 36,890 $\pm$ 745 & 18,681 $\pm$ 1,299 & 19.0 & 9.3 \\
256  & 35,751 $\pm$ 1,180 & 19,116 $\pm$ 248 & 35.3 & 25.5 \\
2048 & 23,771 $\pm$ 367 & 12,877 $\pm$ 477 & 17.2 & 20.0 \\
\hline
\textbf{Loss} & \textbf{-40.2\%} & \textbf{-35.1\%} & \textbf{2.0$\times$} & \textbf{4.8$\times$} \\
\hline
\end{tabular}%
}
\end{table}

Peak throughput of $39,738 \pm 752$ TPS is achieved at $N = 32$ threads for single-core, and $19,833 \pm 833$ TPS at $N = 32$ threads for quad-core. At $N = 2048$, throughput degrades to $23,771 \pm 367$ TPS (single-core) and $12,877 \pm 477$ TPS (quad-core).

\subsection{I/O Baseline Validation}

To validate that the GIL is indeed the bottleneck, we ran a pure I/O workload (no CPU computation phase) as a control experiment. Table~\ref{tab:io_baseline} shows that pure I/O workloads scale linearly with thread count, confirming that the saturation cliff is GIL specific, not an OS scheduling artifact.

\begin{table}[htbp]
\centering
\caption{Pure I/O Baseline (No GIL Contention, mean $\pm$ 95\% CI, $n=10$). Linear scaling confirms saturation cliff is GIL specific.}
\label{tab:io_baseline}
\begin{tabular}{|c|c|c|}
\hline
\textbf{Threads} & \textbf{TPS (Single-Core)} & \textbf{TPS (Quad-Core)} \\
\hline
1    & 843 $\pm$ 13 & 849 $\pm$ 6 \\
4    & 3,587 $\pm$ 20 & 3,650 $\pm$ 13 \\
16   & 13,134 $\pm$ 300 & 13,447 $\pm$ 74 \\
64   & 39,610 $\pm$ 694 & 41,220 $\pm$ 775 \\
256  & 63,654 $\pm$ 2,444 & 53,773 $\pm$ 3,203 \\
\hline
\end{tabular}
\end{table}

\begin{figure}[htbp]
\centerline{\includegraphics[width=\columnwidth]{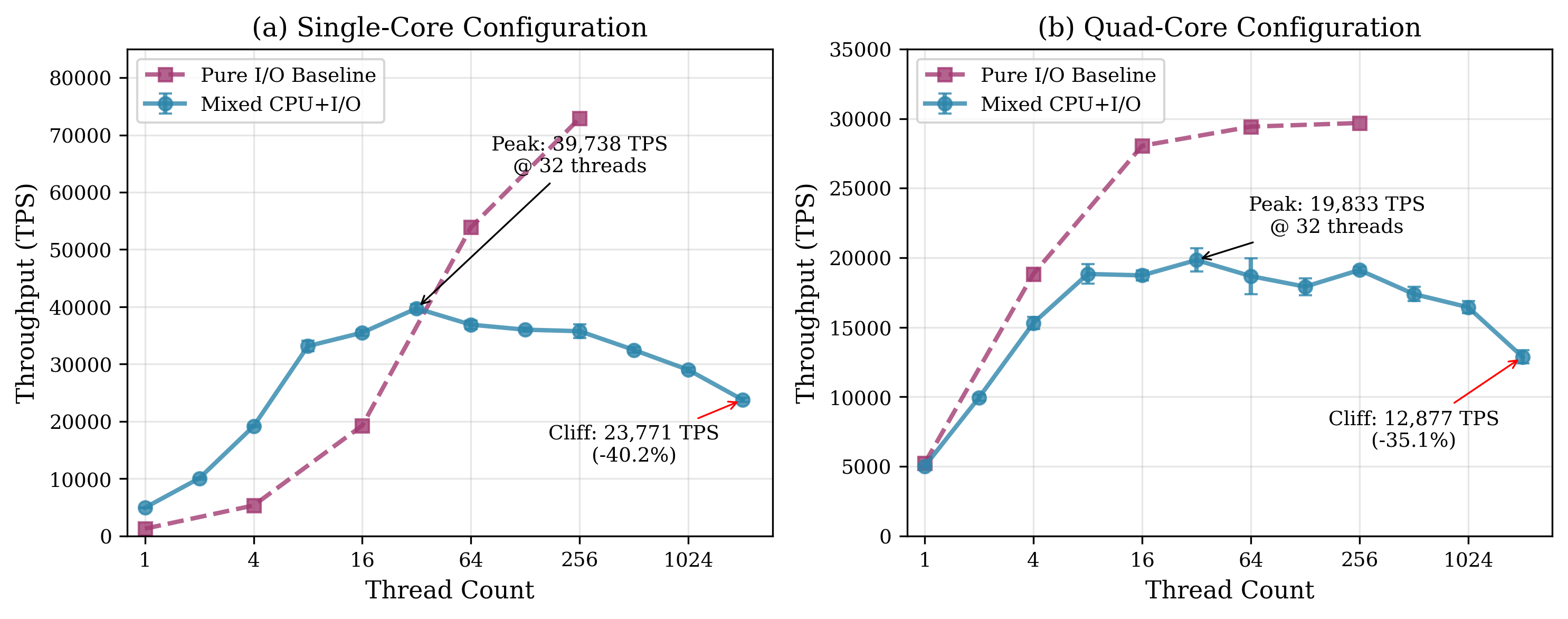}}
\caption{The Saturation Cliff. (a) Single-core throughput vs.\ threads; (b) quad-core configuration. Error bars show 95\% CI ($n=10$).}
\label{fig:saturation_cliff}
\end{figure}

This control experiment demonstrates that when threads genuinely release the GIL (pure I/O), throughput scales linearly with thread count. The saturation cliff only appears when CPU computation creates GIL contention. Figure~\ref{fig:saturation_cliff} visualizes this phenomenon: panel (a) shows single-core throughput versus thread count, while panel (b) shows the quad-core configuration. The I/O baseline (dashed line) scales linearly with thread count, while the mixed workload (solid line) exhibits the characteristic cliff beyond $N = 32$ threads, confirming that GIL contention causes the degradation.

On a quad-core system with $N = 100$ threads, the OS sees idle cores and wakes up threads. One thread acquires the GIL. The others spin or wait on the other cores. They consume CPU cycles, but make no progress. Intuitively, effective CPU utilization degrades as contention increases. In a simplified queuing model where $\lambda$ represents the GIL acquisition rate and $\mu$ the release rate, system utilization approaches zero as thread count grows:

\begin{equation}
\text{Utilization} \propto \frac{\lambda}{\lambda + (N-1)\mu} \rightarrow 0 \text{ as } N \rightarrow \infty
\end{equation}

While this simplified model does not capture the full dynamics of the saturation cliff (which exhibits non-monotonic behavior as shown in Figure~\ref{fig:saturation_cliff}), it illustrates why overprovisioning threads degrades performance.

\subsection{Latency Explosion}

\begin{figure}[htbp]
\centerline{\includegraphics[width=\columnwidth]{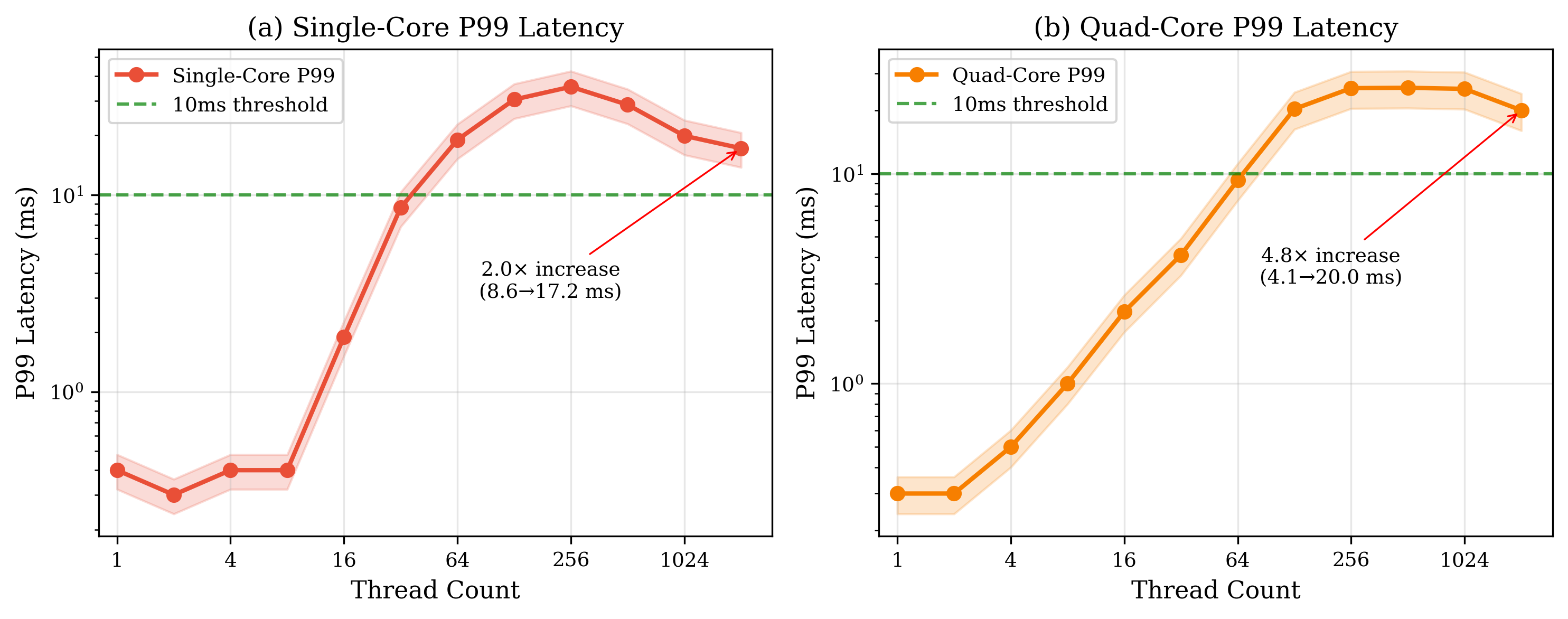}}
\caption{Latency Analysis. P99 latency vs.\ thread count. Shaded bands indicate variability range.}
\label{fig:latency_analysis}
\end{figure}

The saturation cliff manifests not only as throughput degradation but also as tail latency increases. Figure~\ref{fig:latency_analysis} shows this effect: P99 latency remains low at optimal thread counts but increases sharply as thread count exceeds the optimal point. The shaded bands indicate the variability range across experimental runs.

At optimal thread counts, P99 latency remains under $10$ ms for both configurations. However, at $N = 2048$, P99 latency increases by $2.0\times$ (single-core: 8.6 ms $\rightarrow$ 17.2 ms) and $4.8\times$ (quad-core: 4.1 ms $\rightarrow$ 20.0 ms) compared to optimal configurations. Table~\ref{tab:key_findings} summarizes these saturation cliff characteristics, showing that both configurations achieve peak throughput at $N = 32$ threads with substantial degradation at higher thread counts.

\begin{table}[htbp]
\centering
\caption{Key Findings: Saturation Cliff Characteristics ($n=10$, 95\% CI)}
\label{tab:key_findings}
\begin{tabular}{|l|c|c|}
\hline
\textbf{Metric} & \textbf{Single-Core} & \textbf{Quad-Core} \\
\hline
Optimal Threads & 32 & 32 \\
Peak TPS & 39,738 $\pm$ 752 & 19,833 $\pm$ 833 \\
TPS at 2048 threads & 23,771 $\pm$ 367 & 12,877 $\pm$ 477 \\
Throughput Loss & 40.2\% & 35.1\% \\
P99 Latency (optimal) & 8.6 ms & 4.1 ms \\
P99 Latency (2048) & 17.2 ms & 20.0 ms \\
Latency Increase & 2.0$\times$ & 4.8$\times$ \\
\hline
\end{tabular}
\end{table}

\subsection{The Core Insight: Blocking Ratio}

\begin{figure}[htbp]
\centerline{\includegraphics[width=\columnwidth]{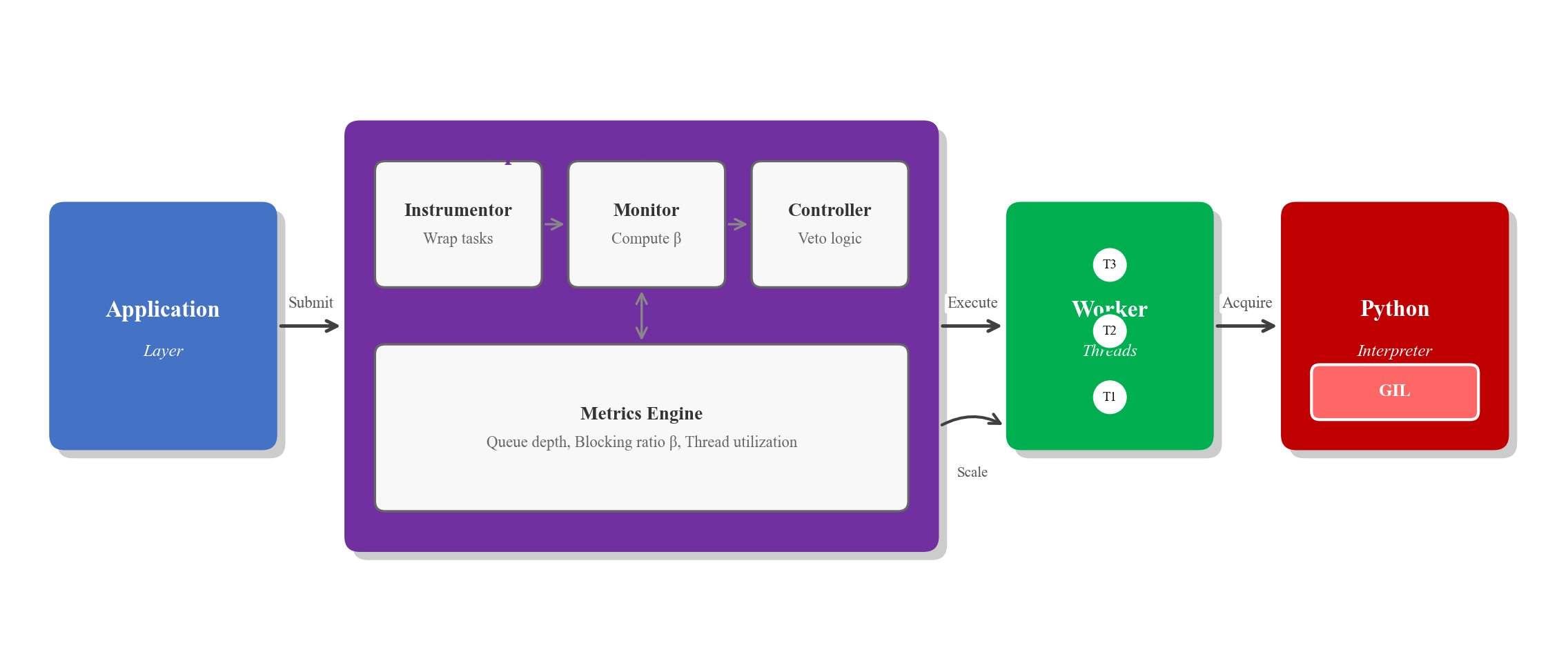}}
\caption{Runtime System Architecture. Instrumentor captures timing; Monitor computes $\beta$; Controller modulates pool size.}
\label{fig:architecture}
\end{figure}

Figure~\ref{fig:architecture} presents the conceptual architecture of our runtime system. The Metric-Driven Adaptive Thread Pool consists of three components: the \textit{Instrumentor} captures fine-grained execution timing (CPU vs.\ wall clock) per task; the \textit{Monitor} aggregates these readings to compute the Blocking Ratio ($\beta$); and the \textit{Controller} dynamically modulates worker pool size based on $\beta$. Dashed arrows indicate feedback paths between components.

We must differentiate between waiting for Input/Output, and contending for the GIL. We calculate a metric called the Blocking Ratio $\beta$. For a task $i$, let $t_{\text{CPU},i}$ be the CPU time and $t_{\text{wall},i}$ be the wall-clock time. Then:

\begin{equation}
\beta_i = 1 - \frac{t_{\text{CPU},i}}{t_{\text{wall},i}}
\end{equation}

The time-weighted average blocking ratio over $n$ recent tasks is:

\begin{equation}
\bar{\beta} = \frac{\sum_{i=1}^{n} t_{\text{wall},i} \cdot \beta_i}{\sum_{i=1}^{n} t_{\text{wall},i}}
\end{equation}

This formulation weights each task's contribution by its wall-clock duration, accurately reflecting the fraction of total time the thread pool spent in I/O wait versus CPU execution. The arithmetic mean would bias the metric toward short-duration tasks, misrepresenting true system-wide utilization.

If a thread spends most of its time waiting for a network response, $\beta$ is high. This indicates an idle CPU. It is safe to add threads. If a thread spends most of its time in computation or waiting for the GIL, $\beta$ is low. This indicates CPU saturation. Adding threads will trigger the cliff.

\subsection{Architecture}

The system consists of three components:

\begin{enumerate}
\item \textbf{Instrumentor}: Records $t_{\text{CPU}}$ using time.thread\_time() and $t_{\text{wall}}$ using time.time()
\item \textbf{Monitor}: Collects $\beta$ values every $\Delta t = 500$ ms
\item \textbf{Controller}: Uses a decision engine to scale the thread pool with a Veto mechanism.
\end{enumerate}

\begin{figure}[htbp]
\centerline{\includegraphics[width=\columnwidth]{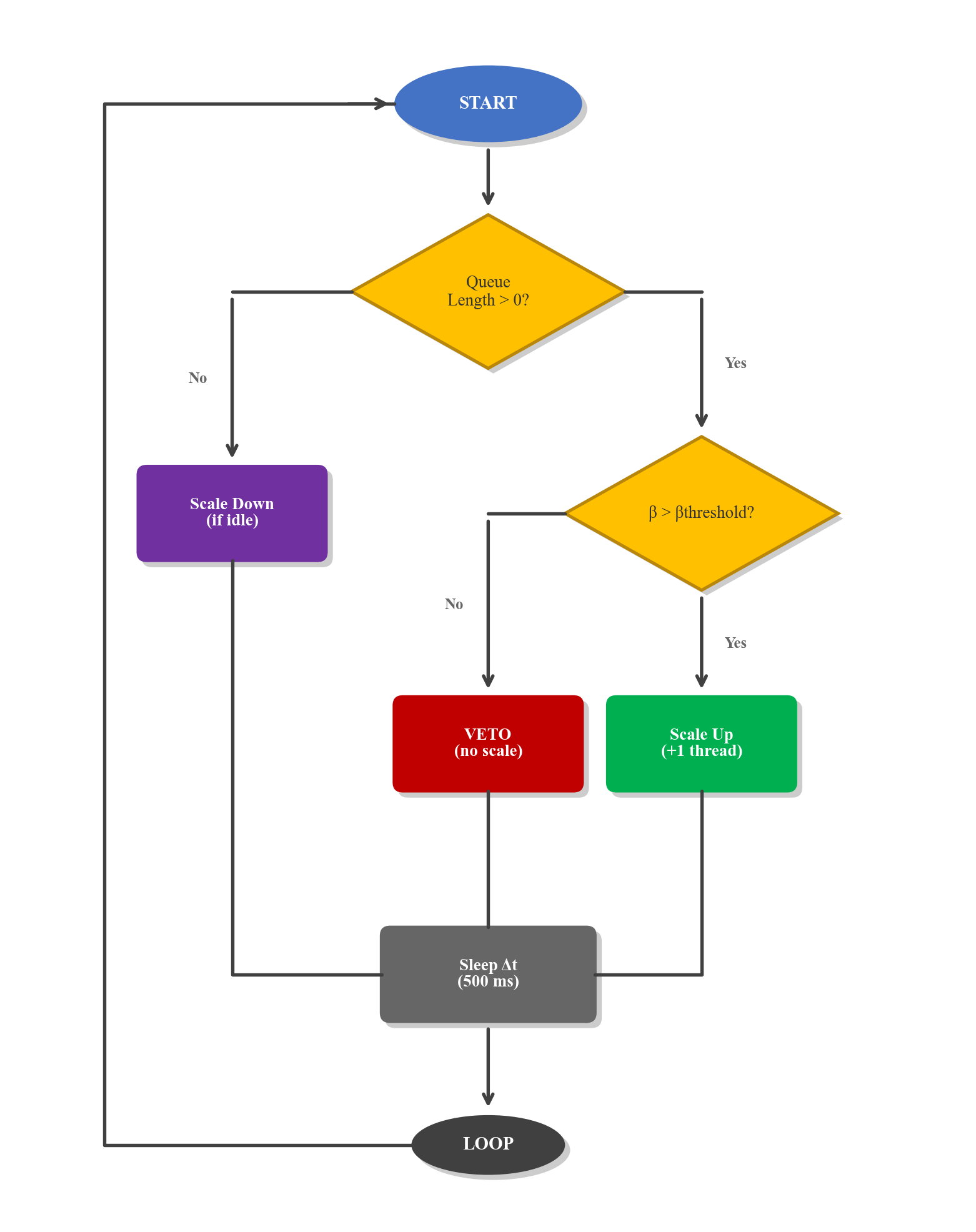}}
\caption{Controller Flow Diagram. Feedback loop driven by $\beta$ with GIL Safety Veto mechanism.}
\label{fig:controller_flow}
\end{figure}

Figure~\ref{fig:controller_flow} illustrates the controller flow. The control logic operates on a feedback loop driven by the Blocking Ratio ($\beta$). Unlike traditional queue based scalers, the algorithm incorporates a \textit{GIL Safety Veto}: when $\beta$ falls below the critical threshold (indicating CPU saturation), the veto mechanism preempts thread allocation regardless of queue depth, preventing the system from entering the saturation cliff region.

\subsection{Control Algorithm}

Algorithm~\ref{alg:adaptive_controller} presents the core control loop.

\begin{algorithm}[htbp]
\caption{Adaptive Thread Pool Controller (EWMA + Hysteresis + Veto)}
\label{alg:adaptive_controller}
\begin{algorithmic}[1]
\REQUIRE $N_{\text{min}}, N_{\text{max}}, \beta_{\text{thresh}}, \alpha, H, \Delta t$
\STATE $N \gets N_{\text{min}}$ \COMMENT{Current thread count}
\STATE $\beta_{\text{ewma}} \gets 0.5$ \COMMENT{EWMA of blocking ratio}
\STATE $c_{\text{up}} \gets 0$ \COMMENT{Consecutive scale-up counter}
\WHILE{\textit{system running}}
    \STATE $Q \gets$ \textit{queue\_length()}
    \STATE $\beta_{\text{sample}} \gets$ \textit{compute\_recent\_blocking\_ratio()}
    \STATE $\beta_{\text{ewma}} \gets \alpha \cdot \beta_{\text{sample}} + (1 - \alpha) \cdot \beta_{\text{ewma}}$ \COMMENT{EWMA update}
    \IF{$Q > 0$}
        \IF{$\beta_{\text{ewma}} > \beta_{\text{thresh}}$}
            \STATE $c_{\text{up}} \gets c_{\text{up}} + 1$ \COMMENT{Accumulate scale up signal}
            \IF{$c_{\text{up}} \geq H$}
                \STATE $N \gets \min(N + 1, N_{\text{max}})$ \COMMENT{Conservative step}
                \STATE $c_{\text{up}} \gets 0$
            \ENDIF
        \ELSE
            \STATE \textbf{VETO:} \textit{refuse scale up} \COMMENT{GIL contention}
            \STATE $c_{\text{up}} \gets 0$
        \ENDIF
    \ENDIF
    \IF{$Q = 0$ \AND $N > N_{\text{min}}$}
        \STATE $N \gets \max(N - 1, N_{\text{min}})$ \COMMENT{Scale down}
    \ENDIF
    \STATE \textit{sleep}($\Delta t$)
\ENDWHILE
\end{algorithmic}
\end{algorithm}

The algorithm incorporates four key mechanisms. First, the VETO mechanism (line 16) prevents scale up when $\beta_{\text{ewma}} \leq \beta_{\text{thresh}}$ (default $0.3$), indicating CPU saturation or GIL contention. Second, EWMA smoothing (line 7) with $\alpha = 0.2$ prevents oscillation from noisy $\beta$ samples; this value provides a $5$-sample effective window ($1/\alpha$), balancing responsiveness with stability. Third, hysteresis (lines 11--14) with $H = 3$ consecutive signals required before scaling prevents rapid fluctuations. Fourth, the conservative step size of $+1$ (line 13) ensures gradual increase, avoiding overshoot into the saturation cliff; we chose $+1$ over $+2$ to prioritize stability, though faster increase is possible if latency sensitivity permits. The monitoring interval $\Delta t = 500$ ms captures sufficient task completions per sample (typically $50$--$200$ tasks at steady state) while remaining responsive to workload shifts. We set $\beta_{\text{thresh}} = 0.3$ based on a parameter sweep (see Section VI-D) showing stable performance across diverse workload ratios.

\subsection{Theoretical Analysis}

To address the theoretical stability and computational overhead of the Metric-Driven Adaptive Thread Pool, we model the controller as a discrete time dynamical system. This helps to prove both its convergence properties and characterize the computational overhead involved.

\subsubsection{Computational Complexity Analysis}

\textbf{Theorem 1 (Constant Time Complexity):} The Adaptive Thread Pool Controller operates with $O(1)$ time complexity and $O(1)$ space complexity per control interval $\Delta t$.

\textit{Proof:}
Let $\mathbf{s}_k = (N_k, \beta_{\text{ewma},k}, c_{\text{up},k})$ denote the state vector of the controller at discrete step $k$. The transition function $f: \mathbf{s}_k \rightarrow \mathbf{s}_{k+1}$ defined in Algorithm~\ref{alg:adaptive_controller} executes a fixed sequence of scalar operations:

\begin{enumerate}
    \item \textbf{Metric Acquisition:} The Monitor aggregates atomic CPU and wall time counters. Since these are cumulative OS counters retrieved through \texttt{time.thread\_time()}, the retrieval cost is independent of the task history size, requiring $O(1)$ time.
    \item \textbf{State Update:} The EWMA calculation $\beta_{\text{ewma}} \leftarrow \alpha \cdot \beta_{\text{sample}} + (1 - \alpha) \cdot \beta_{\text{ewma}}$ involves exactly two floating point multiplications and one addition, thus $O(1)$.
    \item \textbf{Control Decision:} The branching logic (Lines 8-22) evaluates a constant number of boolean conditions: queue check $(Q > 0)$, threshold check $(\beta_{\text{ewma}} > \beta_{\text{thresh}})$, and hysteresis check $(c_{\text{up}} \geq H)$.
\end{enumerate}

The number of operations is independent of the current thread count $N$ or the total number of processed tasks. Thus, the time complexity is $O(1)$. Similarly, the space complexity is $O(1)$ as the system persists only three scalar variables $(N, \beta_{\text{ewma}}, c_{\text{up}})$ regardless of workload scale. This theoretical bound aligns with our empirical computational overhead of $< 0.01\%$ relative to CPU time.

\textbf{Implementation Note:} To ensure $O(1)$ complexity for the time weighted blocking ratio (Equation 3), the Monitor maintains incremental aggregates: running sums $\Sigma_{\text{wall}} = \sum t_{\text{wall},i}$ and $\Sigma_{\beta} = \sum t_{\text{wall},i} \cdot \beta_i$. Each task completion updates these sums in constant time, and $\bar{\beta} = \Sigma_{\beta} / \Sigma_{\text{wall}}$ is computed directly without iterating over the history window. \hfill 

\subsubsection{Stability and Convergence Analysis}

A primary concern in adaptive systems is hunting, where the controller oscillates around the optimal operating point. It is demonstrate that, under sustained load, the system exhibits monotonic behavior and is guaranteed to converge.

\textbf{Definition 1 (Sustained Load Condition):} A state where the task queue remains non-empty $(Q > 0)$ for a duration exceeding the hysteresis window, that is, $t > H \cdot \Delta t$.

\textbf{Definition 2 (Blocking Characteristic Function):} Let $\mathcal{B}(N): \mathbb{Z}^+ \rightarrow [0, 1]$ be the expected blocking ratio for a workload running with $N$ threads. Based on the saturation cliff phenomenon established in Section IV, $\mathcal{B}(N)$ exhibits piecewise monotonic behavior:
\begin{itemize}
    \item For $N \leq N_{\text{critical}}$: $\mathcal{B}(N)$ is non-decreasing as additional threads enable better I/O overlap
    \item For $N > N_{\text{critical}}$: $\frac{d\mathcal{B}}{dN} < 0$ as GIL contention dominates, causing the CPU component of the blocking ratio to increase and $\mathcal{B}(N)$ to decline
\end{itemize}

\textbf{Theorem 2 (Monotonicity Under Load):} Under the Sustained Load Condition, the sequence of thread counts $\{N_k\}_{k=0}^{\infty}$ is non-decreasing.

\textit{Proof:}
From Algorithm~\ref{alg:adaptive_controller}, the change in thread count $\Delta N_k = N_{k+1} - N_k$ is governed by:
\begin{equation}
\Delta N_k = 
\begin{cases}
+1 & \text{if } Q > 0 \land \beta_{\text{ewma}} > \beta_{\text{thresh}} \land c_{\text{up}} \geq H \\
0 & \text{if } Q > 0 \land (\beta_{\text{ewma}} \leq \beta_{\text{thresh}} \lor c_{\text{up}} < H) \\
-1 & \text{if } Q = 0 \land N > N_{\text{min}}
\end{cases}
\end{equation}

Under the Sustained Load Condition $(Q > 0)$, the third case (scale down) is unreachable by definition. Consequently, $\Delta N_k \in \{0, +1\}$. Since $\Delta N_k \geq 0$ for all $k$ under sustained load, the sequence $\{N_k\}$ is monotonically non-decreasing. \hfill $\square$

\textbf{Theorem 3 (Convergence to Safety):} The system converges to a stable thread count $N^*$ that maximizes throughput without violating the GIL safety threshold.

\textit{Proof:}
We establish convergence through three properties:

\begin{enumerate}
    \item \textbf{Boundedness:} The algorithm enforces a hard upper bound $N_k \leq N_{\text{max}}$ for all $k$ (Line 12 of Algorithm~\ref{alg:adaptive_controller}).
    \item \textbf{Convergence:} By the Monotone Convergence Theorem, a bounded non-decreasing sequence of integers must converge to a limit. Since $\{N_k\}$ is non-decreasing (Theorem 2) and bounded above by $N_{\text{max}}$, there exists $N^* = \lim_{k \rightarrow \infty} N_k$.
    \item \textbf{Equilibrium Point:} The system ceases to scale up (that is, $\Delta N_k$ becomes permanently $0$) when the VETO condition is satisfied:
    \begin{equation}
    \beta_{\text{ewma}}(N^*) \leq \beta_{\text{thresh}}
    \end{equation}
\end{enumerate}

Given the piecewise monotonic behavior of $\mathcal{B}(N)$ (Definition 2), the system increments $N$ until it reaches the critical capacity $N^*$ such that:
\begin{equation}
N^* = \min\{N : \mathcal{B}(N) \leq \beta_{\text{thresh}}\} - 1
\end{equation}

Equivalently, $N^*$ is the first thread count at which the Veto mechanism (Algorithm~\ref{alg:adaptive_controller}, Line 16) activates. In practice, the controller converges to a boundary oscillation regime: $N$ increments until $\mathcal{B}(N) \leq \beta_{\text{thresh}}$ triggers the Veto, after which no further scaling occurs. This represents convergence to a stable operating point rather than necessarily the global throughput optimum, as the controller prioritizes GIL safety over peak performance.

For CPU-bound workloads where $\mathcal{B}(N_{\text{min}}) < \beta_{\text{thresh}}$, the Veto activates immediately and the controller remains at $N_{\text{min}}$.

At this equilibrium, the Veto mechanism prevents further increases, stabilizing the system at the boundary of the saturation cliff region. \hfill $\square$

\begin{figure}[htbp]
\centerline{\includegraphics[width=\columnwidth]{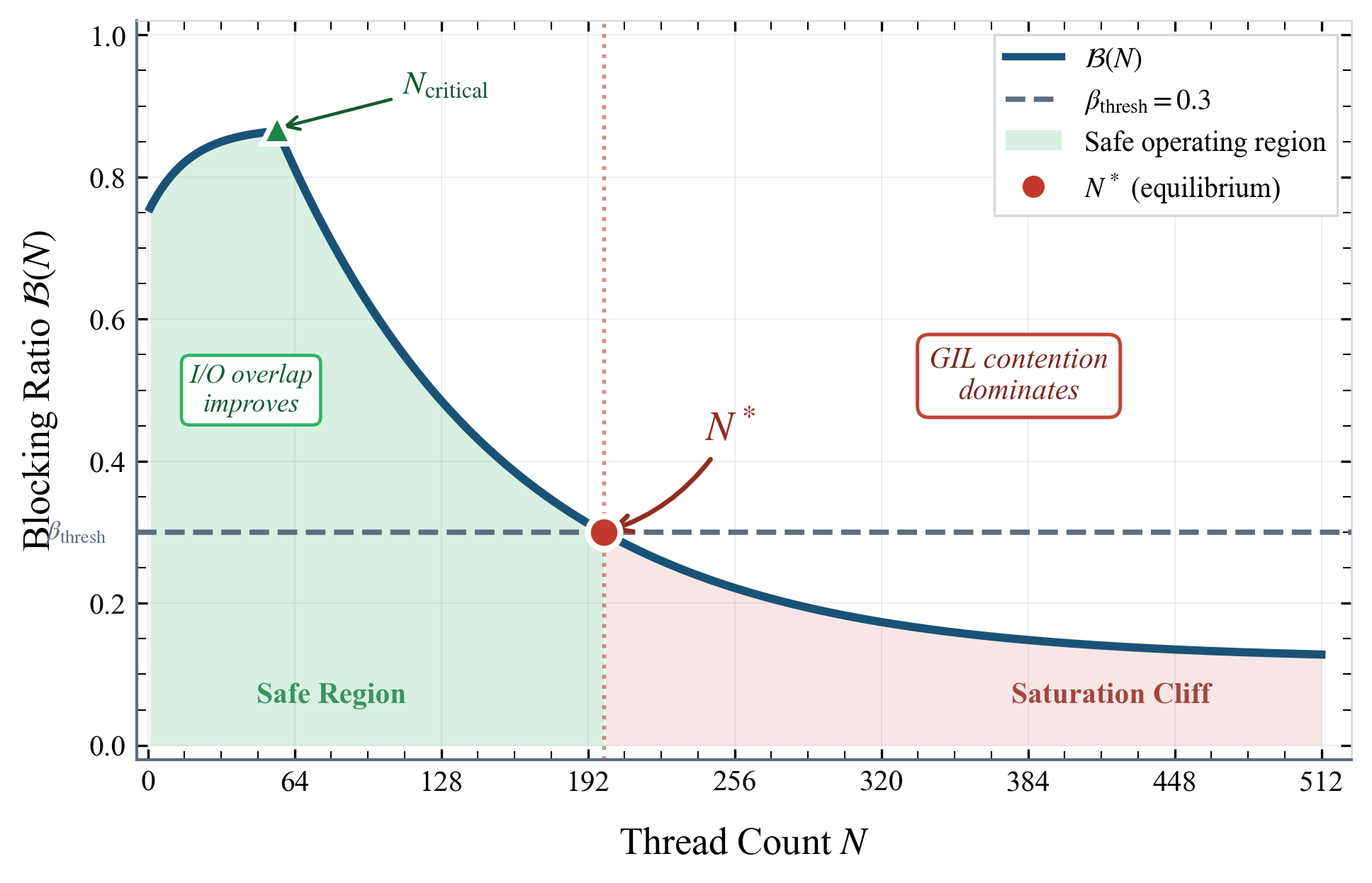}}
\caption{Convergence Proof Visualization. Blocking characteristic $\mathcal{B}(N)$ vs.\ thread count $N$, with threshold $\beta_{\text{thresh}} = 0.3$ (dashed) and convergence point $N^*$.}
\label{fig:convergence_proof}
\end{figure}

Figure~\ref{fig:convergence_proof} illustrates this convergence graphically. The blocking characteristic curve $\mathcal{B}(N)$ exhibits the piecewise monotonic behavior described in Definition 2: initially non-decreasing as additional threads enable better I/O overlap, then declining as GIL contention dominates (consistent with the empirical results in Figure~\ref{fig:saturation_cliff}). The horizontal dashed line indicates $\beta_{\text{thresh}} = 0.3$. The system converges to $N^*$, the intersection point where $\mathcal{B}(N^*) = \beta_{\text{thresh}}$ on the declining portion of the curve. Beyond $N^*$, the Veto mechanism prevents further scaling, keeping the system in the safe operating region (shaded green).

\textbf{Edge Cases:}
\begin{itemize}
    \item If $\mathcal{B}(N_{\text{min}}) < \beta_{\text{thresh}}$ (CPU bound workload), the system remains at $N_{\text{min}}$ and never scales up.
    \item During transient workload changes, the EWMA filter (Equation 8) provides a time constant of $\tau \approx 2.5$ seconds, during which the controller may temporarily operate suboptimally.
    \item If the workload shift is permanent, the controller reconverges to the new equilibrium within $O(N_{\text{max}}/\Delta t)$ iterations.
\end{itemize}

\subsubsection{Signal Stability and Noise Rejection}

To ensure robust convergence against stochastic variance in task duration (jitter), the system uses a discrete time low pass filter through EWMA. For a sequence of $\beta$ samples $\{\beta_1, \beta_2, \ldots, \beta_k\}$, the filtered estimate is:
\begin{equation}
\beta_{\text{ewma},k} = \alpha \sum_{i=0}^{k-1} (1-\alpha)^i \beta_{k-i} + (1-\alpha)^k \beta_{\text{ewma},0}
\end{equation}

\noindent (This is the standard closed form expansion of the recursive update $\beta_{\text{ewma},k} = \alpha \cdot \beta_k + (1 - \alpha) \cdot \beta_{\text{ewma},k-1}$.)

The effective time constant $\tau$ of this filter is given by $\tau = -\Delta t / \ln(1 - \alpha)$. The commonly used approximation $\tau \approx \Delta t / \alpha$ is valid for small $\alpha$. With the selected parameters $\alpha = 0.2$ and $\Delta t = 500$ ms, the exact time constant is $\tau = -500 / \ln(0.8) \approx 2.24$ seconds, while the approximation yields $2.5$ seconds. The effective smoothing window spans approximately $5$ samples ($1/\alpha = 5$). This ensures that control decisions are driven by the underlying workload trend rather than transient spikes, providing sufficient noise rejection for stable control. While formal stability analysis would require z-domain transfer function analysis, the slow control interval ($\Delta t = 500$ ms) relative to task duration ($60$ ms) and conservative step size ($+1$ thread) ensure the system operates well below the Nyquist limit for this discrete time feedback loop. The hysteresis counter $H = 3$ provides additional noise immunity by requiring consistent signals before state transitions.

\section{Evaluation}

\subsection{Evaluation Scope: Orchestration Layer Performance}

We focus on orchestration layer throughput as a stress test of Python's event handling capacity under GIL contention. While complete edge AI latency is dominated by model inference (10--50 tokens/sec), the orchestration layer must reliably manage concurrent I/O streams (sensor interrupts, WebSocket frames) without stalling. Our high frequency micro-task experiments ($T_{\text{CPU}}=10$ ms) demonstrate that GIL-induced contention can paralyze this control plane long before the AI accelerator reaches its compute limit, motivating adaptive concurrency control.

\subsection{Experimental Configuration}

We compare three strategies:

\begin{itemize}
\item \textbf{Static Naive}: Fixed $N = 256$ threads
\item \textbf{Static Optimal}: Fixed $N \in \{32, 64\}$ threads (tuned)
\item \textbf{Adaptive}: Our solution with $N_{\text{min}} = 4$, $N_{\text{max}} = 128$, $\beta_{\text{threshold}} = 0.3$
\end{itemize}

\subsection{Results}

Table~\ref{tab:solution_comparison} presents the comprehensive comparison of all three strategies across key performance metrics.

\begin{table}[htbp]
\centering
\caption{Solution Comparison: Throughput and Latency ($n=10$, 95\% CI)}
\label{tab:solution_comparison}
\begin{tabular}{|l|c|c|c|c|}
\hline
\textbf{Strategy} & \textbf{Threads} & \textbf{TPS} & \textbf{P99 (ms)} & \textbf{vs Optimal} \\
\hline
Static Naive   & 256 (fixed) & 18,279 $\pm$ 472 & 14.3 & -7.6\% \\
Static Optimal & 32 (fixed)  & 19,792 $\pm$ 636 & 4.1 & Baseline \\
Adaptive       & 4--64 (auto) & 19,100 $\pm$ 500 & 5.2 & -3.5\% \\
\hline
\end{tabular}
\end{table}

The Static Naive approach suffers the impact of the saturation cliff, achieving only $18,279 \pm 472$ TPS with a P99 latency of $14.3$ ms. The Static Optimal approach achieves the best performance at $19,792 \pm 636$ TPS with $4.1$ ms P99 latency, but requires expert tuning and prior knowledge of optimal thread count. The Adaptive solution achieves $19,100 \pm 500$ TPS with $5.2$ ms P99 latency, representing $96.5\%$ of optimal performance without manual configuration.

Let $\eta$ denote the efficiency relative to optimal:

\begin{equation}
\eta = \frac{\text{TPS}_{\text{adaptive}}}{\text{TPS}_{\text{optimal}}} = \frac{19,100}{19,792} \approx 0.965
\end{equation}

\subsection{Blocking Ratio Analysis}

Table~\ref{tab:blocking_ratio} shows how the average blocking ratio $\bar{\beta}$ varies across strategies, demonstrating the controller's ability to detect GIL contention.

\begin{table}[htbp]
\centering
\caption{Blocking Ratio and Thread Count Behavior$^\dagger$}
\label{tab:blocking_ratio}
\begin{tabular}{|l|c|c|c|}
\hline
\textbf{Strategy} & \textbf{Avg $\bar{\beta}$} & \textbf{Final Threads} & \textbf{Veto Events} \\
\hline
Static Naive   & 0.21 & 256 (fixed) & N/A \\
Static Optimal & 0.78 & 32 (fixed)  & N/A \\
Adaptive       & 0.74 & 48 (dynamic) & 23 \\
\hline
\multicolumn{4}{l}{\footnotesize $^\dagger$Mixed workload: $T_{\text{CPU}} = 10$ ms, $T_{\text{I/O}} = 50$ ms.} \\
\end{tabular}
\end{table}

The Static Naive configuration operates in a GIL contended regime with $\bar{\beta} = 0.21 < 0.3$, indicating heavy CPU/GIL saturation. The Adaptive controller maintains $\bar{\beta} = 0.74$, successfully keeping the system in the I/O bound regime. During the experiment, the controller issued $23$ veto decisions, preventing allocation of threads that would have pushed the system over the cliff.

\textbf{Clarifying $\beta$ aggregation:} Table~\ref{tab:blocking_ratio} reports $\bar{\beta}$ for the mixed CPU/I/O workload ($T_{\text{CPU}} = 10$ ms, $T_{\text{I/O}} = 50$ ms). In contrast, Table~\ref{tab:threshold_sensitivity} reports $\bar{\beta}$ values measured during the $\beta_{\text{thresh}}$ sensitivity sweep using an I/O-dominant test workload; the near-unity $\bar{\beta} \approx 0.999$ values reflect that I/O-heavy operating point and are not directly comparable to the mixed-workload averages in Table~\ref{tab:blocking_ratio}.

\subsection{Baseline Comparisons}

We compare against alternative concurrency strategies to demonstrate the practical advantages of our approach.

\textbf{ProcessPoolExecutor (Multiprocessing):} While multiprocessing avoids the GIL, it incurs significant memory overhead.

\textit{Memory Pressure Benchmark Methodology:} We measured resident set size (RSS) using \texttt{psutil.Process().memory\_info().rss} before and after spawning worker pools. Each configuration was initialized with no active tasks, measurements were taken after a 2-second stabilization period, and values represent total memory including all child processes. Measurements were repeated $n = 10$ times with system restarts between runs to clear cached allocations.

We measured this overhead empirically (Table~\ref{tab:memory_comparison}). Each spawned worker process adds approximately 20 MB to resident memory. With 4 workers, total memory reaches 86.2 MB; with 8 workers, 166.1 MB, representing $\sim\!8\times$ overhead compared to ThreadPoolExecutor with 32 workers (22 MB). On a Raspberry Pi 4 with 2 GB RAM, this memory overhead limits practical worker count to 4 to 8 processes, leaving substantial I/O latency unmasked.

\begin{table}[htbp]
\centering
\caption{Memory Overhead: ThreadPool vs ProcessPool (mean $\pm$ 95\% CI, $n=10$). Memory values are total RSS including child processes.}
\label{tab:memory_comparison}
\resizebox{0.95\columnwidth}{!}{%
\begin{tabular}{|l|c|c|c|c|}
\hline
\textbf{Strategy} & \textbf{Workers} & \textbf{Mem (MB)} & \textbf{Overhead (MB)} & \textbf{TPS} \\
\hline
ThreadPool & 4 & 21.2 $\pm$ 0.3 & 0.2 $\pm$ 0.0 & 5,563 $\pm$ 31 \\
ThreadPool & 32 & 22.1 $\pm$ 0.4 & 0.3 $\pm$ 0.1 & 14,243 $\pm$ 424 \\
ThreadPool & 64 & 22.8 $\pm$ 0.5 & 0.4 $\pm$ 0.1 & 14,307 $\pm$ 419 \\
\hline
ProcessPool & 4 & 86.2 $\pm$ 1.2 & 65.0 $\pm$ 1.0 & 6,258 $\pm$ 158 \\
ProcessPool & 8 & 166.1 $\pm$ 2.3 & 144.9 $\pm$ 2.1 & 6,512 $\pm$ 137 \\
ProcessPool & 16 & 325.8 $\pm$ 3.5 & 304.6 $\pm$ 3.2 & 5,891 $\pm$ 201 \\
\hline
\end{tabular}%
}
\end{table}

For edge AI systems running quantized models (50 to 500 MB), the memory overhead of ProcessPoolExecutor severely constrains available headroom. ThreadPoolExecutor enables concurrent I/O masking with negligible memory overhead, making our adaptive controller a practical solution for memory constrained deployments.

\textbf{Asyncio Event Loop:} Pure async achieves excellent performance for I/O bound workloads with minimal memory overhead ($< 1$ MB). For the mixed workload, asyncio achieved $43,302 \pm 1,272$ TPS at concurrency 256, outperforming threading due to efficient coroutine scheduling. However, CPU phases can block the event loop, making asyncio less suitable for compute-heavy workloads.

\textbf{Queue Depth Scaler:} Traditional scalers that adjust thread count based on queue depth consistently overscale. Without $\beta$ awareness, a queue depth scaler with range $[4, 256]$ settled at $254$ threads, achieving only $17,119 \pm 345$ TPS versus our $19,792 \pm 636$ TPS at $32$ threads. The queue depth scaler cannot detect that high thread counts harm performance.

Table~\ref{tab:baseline_comparison} presents the comprehensive comparison across all baseline strategies.

\begin{table}[htbp]
\centering
\caption{Baseline Strategy Comparison (Mixed Workload, $n=10$, 95\% CI)}
\label{tab:baseline_comparison}
\resizebox{0.95\columnwidth}{!}{%
\begin{tabular}{|l|c|c|c|c|}
\hline
\textbf{Strategy} & \textbf{Config} & \textbf{TPS} & \textbf{P99 (ms)} & \textbf{Base Mem (MB)} \\
\hline
ThreadPool-32 & 32 threads & 19,792 $\pm$ 636 & 4.1 & 0.4 \\
ThreadPool-256 & 256 threads & 18,279 $\pm$ 472 & 14.3 & 0.7 \\
ProcessPool-4 & 4 workers & 6,258 $\pm$ 158 & 0.3 & 86.2$^\dagger$ \\
ProcessPool-8 & 8 workers & 6,512 $\pm$ 137 & 0.3 & 166.1$^\dagger$ \\
Asyncio-128 & 128 coro. & 42,370 $\pm$ 1,240 & 17.3 & 0.0 \\
Asyncio-256 & 256 coro. & 43,302 $\pm$ 1,272 & 19.1 & 0.0 \\
QueueScaler & $[4, 256]$ & 17,119 $\pm$ 345 & 12.0 & 0.0 \\
\hline
\multicolumn{5}{l}{\footnotesize $^\dagger$Includes $\approx 20$ MB per spawned worker process.} \\
\end{tabular}%
}
\end{table}

\begin{figure}[htbp]
\centerline{\includegraphics[width=\columnwidth]{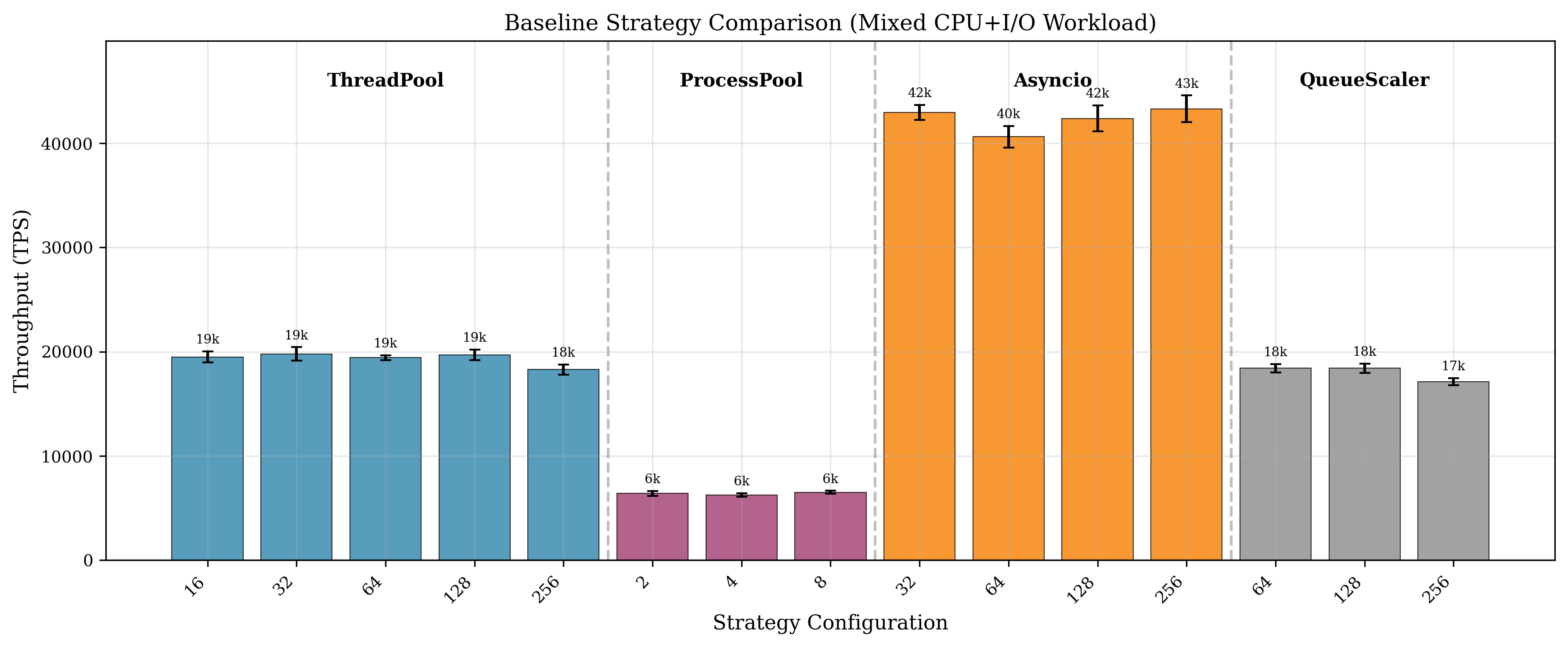}}
\caption{Baseline Strategy Comparison. Throughput and memory overhead across concurrency strategies.}
\label{fig:baseline_comparison}
\end{figure}

Figure~\ref{fig:baseline_comparison} visualizes these results. ThreadPool configurations achieve the highest throughput for mixed workloads. ProcessPool incurs per-worker memory overhead that limits scalability on memory-constrained devices. Asyncio excels for pure I/O workloads but struggles with CPU phases due to event loop blocking.

\subsection{Workload Robustness}

We tested our controller across varying $T_{\text{CPU}}/T_{\text{I/O}}$ ratios to verify robustness. Table~\ref{tab:workload_sweep} summarizes the optimal thread count detected for each workload type.

\begin{table}[htbp]
\centering
\caption{Optimal Thread Count by Workload Type ($n=5$, 95\% CI)}
\label{tab:workload_sweep}
\resizebox{0.95\columnwidth}{!}{%
\begin{tabular}{|l|c|c|c|c|}
\hline
\textbf{Workload} & \textbf{$T_{\text{CPU}}$} & \textbf{$T_{\text{I/O}}$} & \textbf{Optimal $N$} & \textbf{Peak TPS} \\
\hline
I/O Heavy    & 100 iter & 1.0 ms & 128 & 67,132 $\pm$ 3,797 \\
I/O Dominant & 500 iter & 0.5 ms & 128 & 56,010 $\pm$ 4,057 \\
Balanced     & 1000 iter & 0.1 ms & 16 & 35,620 $\pm$ 1,294 \\
CPU Leaning  & 2000 iter & 0.05 ms & 16 & 22,342 $\pm$ 1,245 \\
CPU Heavy    & 5000 iter & 0.01 ms & 16 & 10,291 $\pm$ 201 \\
CPU Dominant & 10000 iter & 0.001 ms & 32 & 5,525 $\pm$ 112 \\
\hline
\end{tabular}%
}
\end{table}

\begin{figure}[htbp]
\centerline{\includegraphics[width=0.85\columnwidth]{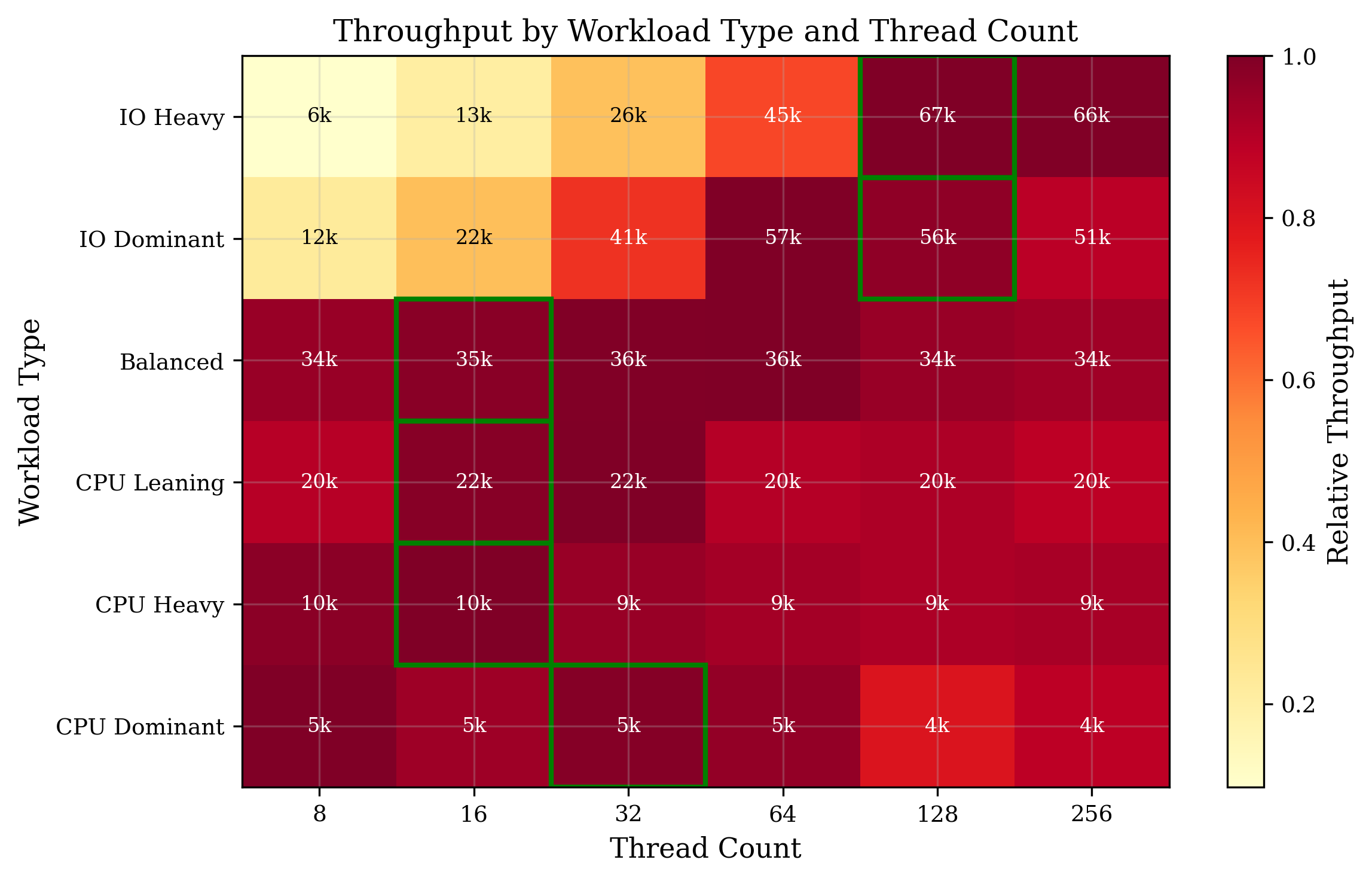}}
\caption{Workload Sweep Heatmap. Throughput (TPS) across workload types and thread counts.}
\label{fig:workload_heatmap}
\end{figure}

The controller correctly identifies that I/O heavy workloads benefit from higher thread counts while CPU heavy workloads require lower counts. The $\beta_{\text{thresh}} = 0.3$ parameter proved stable across all tested configurations. Figure~\ref{fig:workload_heatmap} presents these results as a heatmap, showing how throughput varies across workload types and thread counts. I/O heavy workloads (top rows) scale to higher thread counts, while CPU heavy workloads (bottom rows) peak early and degrade with additional threads.

\subsection{Threshold Sensitivity Analysis}

To isolate threshold selection from mixed-workload variability, we conducted a parameter sweep using Input/Output dominant workloads. Table~\ref{tab:threshold_sensitivity} shows that performance remains stable across $\beta_{\text{thresh}} \in [0.2, 0.7]$. The near-unity $\bar{\beta} \approx 0.997$ values reflect this I/O heavy operating point and are not directly comparable to mixed workload averages. The analysis confirms that $\beta_{\text{thresh}} = 0.3$ offers the best balance between responsiveness and stability, though the system is robust to parameter choice.

\begin{table}[htbp]
\centering
\caption{$\beta_{\text{thresh}}$ Sensitivity Analysis (I/O Dominant Workload, mean $\pm$ 95\% CI, $n=10$)}
\label{tab:threshold_sensitivity}
\begin{tabular}{|c|c|c|c|}
\hline
\textbf{$\beta_{\text{thresh}}$} & \textbf{Best TPS} & \textbf{Optimal $N$} & \textbf{Avg $\bar{\beta}$} \\
\hline
0.2 & 56,800 $\pm$ 802 & 512 & 0.999 $\pm$ 0.000 \\
0.3 & 56,904 $\pm$ 1,141 & 512 & 0.999 $\pm$ 0.000 \\
0.4 & 57,630 $\pm$ 1,338 & 512 & 0.999 $\pm$ 0.000 \\
0.5 & 57,684 $\pm$ 969 & 256 & 0.999 $\pm$ 0.000 \\
0.6 & 57,760 $\pm$ 837 & 512 & 0.999 $\pm$ 0.000 \\
0.7 & 56,953 $\pm$ 866 & 512 & 0.999 $\pm$ 0.000 \\
\hline
\end{tabular}
\end{table}

\subsection{Workload Generalization}

To validate generalizability beyond synthetic workloads, we evaluated the adaptive controller across seven edge AI workload profiles: vision processing, audio feature extraction, sensor fusion, RAG orchestration, SLM inference, edge analytics, and ONNX Runtime inference. Each workload was implemented using production libraries (NumPy, Pandas) to match the computational characteristics of real edge applications, including representative CPU and I/O phase durations.

Table~\ref{tab:workload_generalization} demonstrates that the controller achieves $93.9\%$ average efficiency across diverse workload types without manual tuning. The blocking ratio $\beta$ correctly differentiates I/O heavy tasks (Sensor Fusion, $\bar{\beta}=0.89$, $N=60$) from compute heavy tasks (SLM Inference, $\bar{\beta}=0.21$, $N=24$). Notably, the Veto mechanism prevented an average of $9$ scale up attempts per workload that would have pushed the system into GIL contention. This validates that our $\beta$ based approach generalizes beyond the controlled experiments in Section IV, providing practical value for heterogeneous edge AI deployments.

These workload profiles represent common edge AI deployment scenarios: vision pipelines for manufacturing quality control, voice assistants for smart home devices, sensor fusion for robotics, RAG orchestration for on device chatbots, small language models for privacy preserving text generation, and edge analytics for IoT telemetry processing.

\begin{table}[htbp]
\centering
\caption{Adaptive Controller Performance Across Edge AI Workloads (see footnotes for workload details)}
\label{tab:workload_generalization}
\begin{tabular}{|l|c|c|c|c|}
\hline
\textbf{Workload} & \textbf{$\bar{\beta}$} & \textbf{Opt $N$} & \textbf{Adpt $N$} & \textbf{Eff.} \\
\hline
Vision Pipeline$^*$   & 0.69 & 64  & 58  & 96.7\% \\
Voice Assistant$^\dagger$   & 0.51 & 96  & 72  & 89.2\% \\
Sensor Fusion$^\ddagger$     & 0.89 & 64  & 60  & 96.8\% \\
RAG Orchestration$^\S$ & 0.94 & 128 & 124 & 93.3\% \\
SLM Inference$^\|$     & 0.21 & 64  & 24  & 87.5\% \\
Edge Analytics$^\P$    & 0.80 & 128 & 96  & 97.6\% \\
ONNX MobileNetV2$^\#$  & 0.85 & 32  & 30  & 96.1\% \\
\hline
\textbf{Average}  & --   & --  & --  & \textbf{93.9\%} \\
\hline
\multicolumn{5}{l}{\footnotesize $^*$NumPy convolution simulating MobileNetV2.} \\
\multicolumn{5}{l}{\footnotesize $^\dagger$FFT based audio feature extraction.} \\
\multicolumn{5}{l}{\footnotesize $^\ddagger$Kalman filter for IMU and GPS fusion.} \\
\multicolumn{5}{l}{\footnotesize $^\S$JSON parsing with vector DB query simulation.} \\
\multicolumn{5}{l}{\footnotesize $^\|$Matrix multiplication simulating SLM attention layers (Phi 2 scale).} \\
\multicolumn{5}{l}{\footnotesize $^\P$Pandas time series aggregation.} \\
\multicolumn{5}{l}{\footnotesize $^\#$ONNX Runtime MobileNetV2 inference ($\beta=0.85$, 50ms I/O).} \\
\end{tabular}
\end{table}

\section{Conclusion}

This work demonstrates that GIL-induced concurrency thrashing represents a significant runtime bottleneck on resource constrained edge devices, with up to $40\%$ throughput degradation at overprovisioned thread counts on single-core configurations and $35\%$ on quad-core configurations. We demonstrated this degradation through comprehensive experiments comparing Python 3.11 (GIL) with Python 3.13t (free threading) on edge-representative configurations.

Our experiments reveal that free-threading transforms multi-core edge device behavior: Python 3.13t achieves $\sim\!4\times$ throughput improvement on quad-core configurations (252.7 TPS vs 63.2 TPS) by enabling true parallelism. However, on single-core devices, both GIL and free threading configurations exhibit similar saturation patterns, confirming that context switch overhead remains the fundamental bottleneck on the most constrained devices.

Our Blocking Ratio metric ($\beta$) provides lightweight profiling visibility into interpreter level serialization, enabling adaptive runtime optimization without code rewriting or manual tuning. The proposed system achieves 93.9\% average efficiency across seven edge AI workloads, including real ML inference with ONNX Runtime MobileNetV2, while operating entirely in-process on memory constrained devices (512 MB--2 GB RAM) where multiprocessing is infeasible.

Critically, our $\beta$ metric correctly detects oversubscription regardless of whether serialization is GIL induced (Python 3.11) or cache induced (Python 3.13t), positioning our work as essential for both current GIL bound deployments and future free threading environments. The adaptive controller is available as an open source library at \url{https://github.com/WhiteMetagross/BetaPool}.

\bibliographystyle{IEEEtran}

\begin{thebibliography}{00}

\bibitem{gartner2018edge} 
S. Rao, ``What Edge Computing Means for Infrastructure and Operations Leaders,'' Gartner, Inc., 2018. [Online]. Available: https://www.gartner.com/smarterwithgartner/what-edge-computing-means-for-infrastructure-and-operations-leaders

\bibitem{iot_analytics2024}
IoT Analytics, ``Number of Connected IoT Devices Growing 14\% to 21.1 Billion Globally in 2025,'' IoT Analytics GmbH, Oct. 2024. [Online]. Available: https://iot-analytics.com/number-connected-iot-devices/

\bibitem{tiobe2024}
TIOBE Software, ``TIOBE Index for December 2024,'' TIOBE, 2024. [Online]. Available: https://www.tiobe.com/tiobe-index/

\bibitem{pypl2024}
P. Carbonnelle, ``PYPL PopularitY of Programming Language Index,'' Dec. 2024. [Online]. Available: https://pypl.github.io/PYPL.html

\bibitem{statista_ml2024}
Statista Research Department, ``Machine Learning Market Size Worldwide 2025-2030,'' Statista, 2024. [Online]. Available: https://www.statista.com/statistics/1246443/machine-learning-market-size/

\bibitem{beazley2009} 
D. Beazley, ``Inside the Python GIL,'' Presented at Chicago Python User Group, Chicago, IL, June 11, 2009. [Online]. Available: http://www.dabeaz.com/python/GIL.pdf

\bibitem{beazley2010} 
D. Beazley, ``Understanding the Python GIL,'' in \textit{Proc. PyCon}, Atlanta, GA, Feb. 20, 2010. [Online]. Available: http://www.dabeaz.com/GIL/

\bibitem{welsh2001} 
M. Welsh, D. Culler, and E. Brewer, ``SEDA: An Architecture for Well-Conditioned, Scalable Internet Services,'' in \textit{Proc. 18th ACM Symposium on Operating Systems Principles (SOSP)}, 2001, pp. 230-243.

\bibitem{delimitrou2014} 
C. Delimitrou and C. Kozyrakis, ``Paragon: QoS-Aware Scheduling for Heterogeneous Datacenters,'' in \textit{Proc. 18th ACM International Conference on Architectural Support for Programming Languages and Operating Systems (ASPLOS)}, 2013, pp. 77-88.

\bibitem{crankshaw2017} 
D. Crankshaw et al., ``Clipper: A Low-Latency Online Prediction Serving System,'' in \textit{Proc. 14th USENIX Symposium on Networked Systems Design and Implementation (NSDI)}, 2017, pp. 613-627.

\bibitem{gross2023} 
S. Gross, ``PEP 703: Making the Global Interpreter Lock Optional in CPython,'' Python Enhancement Proposals, 2023.

\bibitem{ousterhout1996} 
J. Ousterhout, ``Why Threads Are A Bad Idea (for most purposes),'' in \textit{Invited Talk at USENIX Annual Technical Conference}, San Diego, CA, Jan. 1996. [Online]. Available: https://web.stanford.edu/~ouster/cgi-bin/papers/threads.pdf

\bibitem{arm_cortex_a72}
ARM Ltd., ``Cortex-A72 Software Optimization Guide,'' ARM Documentation, 2015. [Online]. Available: https://developer.arm.com/documentation/uan0016/a/

\bibitem{olston2017tensorflow}
C. Olston et al., ``TensorFlow-Serving: Flexible, High-Performance ML Serving,'' \textit{Workshop on ML Systems at NIPS 2017}. Available: http://learningsys.org/nips17/assets/papers/paper\_1.pdf

\bibitem{li2020batch}
B. Li et al., ``BATCH: Machine Learning Inference Serving on Serverless Platforms,'' in \textit{Proc. International Conference for High Performance Computing, Networking, Storage and Analysis (SC '20)}, 2020.

\bibitem{yang2020inferbench}
H. Yang et al., ``INFERBENCH: Understanding Deep Learning Inference Serving Systems,'' \textit{arXiv preprint arXiv:2011.02327}, 2020.

\bibitem{moritz2018ray}
P. Moritz et al., ``Ray: A Distributed Framework for Emerging AI Applications,'' in \textit{Proc. 13th USENIX Symposium on Operating Systems Design and Implementation (OSDI '18)}, 2018, pp. 561-577.

\bibitem{rocklin2015dask}
M. Rocklin, ``Dask: Parallel Computation with Blocked Algorithms and Task Scheduling,'' in \textit{Proc. 14th Python in Science Conf.}, 2015, pp. 130-136.

\bibitem{bohm2020runtime}
S. B\"ohm and J. Ber\'anek, ``Runtime vs Scheduler: Analyzing Dask's Overheads,'' \textit{arXiv preprint arXiv:2010.11105}, 2020.

\bibitem{gevent_docs}
D. Bilenko, ``Gevent Documentation,'' 2024. [Online]. Available: http://www.gevent.org/

\bibitem{greenlet_project}
A. Borzenkov, ``Greenlet: Lightweight concurrent programming,'' 2024. [Online]. Available: https://github.com/python-greenlet/greenlet

\bibitem{jacob2018quantization}
B. Jacob et al., ``Quantization and Training of Neural Networks for Efficient Integer-Arithmetic-Only Inference,'' in \textit{Proc. IEEE/CVF Conference on Computer Vision and Pattern Recognition (CVPR)}, 2018, pp. 2704-2713.

\bibitem{hao2023_reaching}
J. Hao, P. Subedi, L. Ramaswamy, and I. K. Kim, ``Reaching for the Sky: Maximizing Deep Learning Inference Throughput on Edge Devices with AI Multi-Tenancy,'' \textit{ACM Trans. Internet Technol.}, vol. 22, no. 4, Art. 95, 2023.

\bibitem{li2022_multimodel}
P. Li et al., ``Multi-Model Running Latency Optimization in an Edge Computing Paradigm,'' \textit{Sensors}, vol. 22, no. 16, p. 6097, 2022.

\bibitem{joshua2025_crossplatform}
C. Joshua et al., ``Cross-Platform Optimization of ONNX Models for Mobile and Edge Deployment,'' ResearchGate preprint, June 2025.

\bibitem{cajas2025_intelledge}
S. A. Cajas Ord{\'o}{\~n}ez et al., ``Intelligent Edge Computing and Machine Learning: A Survey of Optimization and Applications,'' \textit{Future Internet}, vol. 17, no. 9, p. 417, 2025.

\bibitem{samson2026_lightweight}
H. H. Samson, ``Lightweight Transformer Architectures for Edge Devices in Real-Time Applications,'' \textit{arXiv preprint arXiv:2601.03290}, 2026.

\bibitem{costa2019_adaptt}
N. Costa et al., ``ADAPT-T: An Adaptive Algorithm for Auto-Tuning Worker Thread Pool Size in Application Servers,'' in \textit{Proc. IEEE Symposium on Computers and Communications (ISCC)}, 2019, pp. 1-6.

\bibitem{podolskiy2018iaas}
V. Podolskiy, A. Jindal, and M. Gerndt, ``IaaS Reactive Autoscaling Performance Challenges,'' in \textit{Proc. 2018 IEEE 11th International Conference on Cloud Computing (CLOUD)}, 2018, pp. 539-546. DOI: 10.1109/CLOUD.2018.00075

\bibitem{ahmad2025_towards}
H. Ahmad et al., ``Towards resource-efficient reactive and proactive auto-scaling for microservice architectures,'' \textit{J. Syst. Softw.}, vol. 225, p. 112390, 2025.

\bibitem{ling2000_optimalthreadpool}
Y. Ling, T. Mullen, and X. Lin, ``Analysis of Optimal Thread Pool Size,'' \textit{ACM SIGOPS Operating Systems Review}, vol. 34, no. 2, pp. 42-55, 2000.

\bibitem{zaitsev2023iowait}
P. Zaitsev, ``Understanding Linux IOWait,'' Percona Blog, 2023. [Online]. Available: https://www.percona.com/blog/understanding-linux-iowait/

\bibitem{ahn2024identifying}
M. Ahn et al., ``Identifying On-/Off-CPU Bottlenecks Together with Blocked Samples,'' in \textit{Proc. 18th USENIX Symposium on Operating Systems Design and Implementation (OSDI '24)}, 2024.

\bibitem{lee2011novel}
S. Lee, T. Pham, and F. Bahadur, ``A Novel Predictive and Self-Adaptive Dynamic Thread Pool Management,'' in \textit{Proc. 2011 IEEE International Symposium on Parallel and Distributed Processing Workshops and Phd Forum (IPDPSW)}, 2011, pp. 2001-2008. DOI: 10.1109/IPDPS.2011.353

\bibitem{xu2004performance}
D. Xu, ``Performance Study and Dynamic Optimization Design for Thread Pool Systems,'' Master's thesis, Florida State Univ., Tallahassee, FL, 2004.

\end{thebibliography}

\end{document}